\documentclass{article}[9pt]
\usepackage{amssymb}
\usepackage{textgreek}
\usepackage{graphicx}
\usepackage{newtxtext,newtxmath}
\usepackage[margin=3cm]{geometry}
\def\be{\begin{equation}} 
\def\ee{\end{equation}} 
\def\ba{\begin{eqnarray}} 
\def\ea{\end{eqnarray}} 
\def\bc{\begin{center}} 
\def\ec{\end{center}} 
\usepackage[
    safeinputenc,
    backend=biber,
    style=numeric-comp,
    sorting=none,
    natbib=false,
    url=false,
    doi=false,
    eprint=false,
    maxbibnames=999,
    giveninits=true
]{biblatex}
\renewbibmacro{in:}{}
\usepackage{xpatch}
\xpatchbibdriver{book}{%
\newunit\newblock
\printfield{edition}%
}
{%
\setunit{\addcomma\space}\newblock
\printfield{edition}%
}{}{}
\renewcommand\labelnamepunct{\addcomma\space}
\DeclareFieldFormat[misc]{title}{#1}
\DefineBibliographyStrings{english}{%
  page             = {\ifbibliography{}{\adddot}},
  pages            = {\ifbibliography{}{\adddot}},
}

\renewbibmacro*{volume+number+eid}{%
  \setunit*{\addcomma\space}
  \printfield{volume}%
  \setunit*{\addcomma\space}
  \printfield{number}%
  \setunit{\addcomma\space}%
  \printfield{eid}}
\DeclareFieldFormat[article]{volume}{\mkbibbold{#1}}

\renewbibmacro*{journal+issuetitle}{%
  \usebibmacro{journal}%
  \setunit*{\addspace}%
  \iffieldundef{series}
    {}
    {\newunit
     \printfield{series}%
     \setunit{\addspace}}%
  \usebibmacro{volume+number+eid}%
  \newunit} 
  
\DeclareBibliographyDriver{article}{%
  \usebibmacro{bibindex}%
  \usebibmacro{begentry}%
  \usebibmacro{author/translator+others}%
  \setunit{\labelnamepunct}\newblock
  \usebibmacro{title}%
  \newunit
  \printlist{language}%
  \newunit\newblock
  \usebibmacro{byauthor}%
  \newunit\newblock
  \usebibmacro{bytranslator+others}%
  \newunit\newblock
  \printfield{version}%
  \newunit\newblock
  \usebibmacro{journal+issuetitle}%
  \newunit
  \usebibmacro{byeditor+others}%
  \newunit
  \usebibmacro{note+pages}%
  \setunit{\addspace}
  \usebibmacro{issue+date}
  \setunit{\addcolon\space}
  \usebibmacro{issue}
  \newunit\newblock
  \iftoggle{bbx:isbn}
    {\printfield{issn}}
    {}%
  \newunit\newblock
  \usebibmacro{doi+eprint+url}%
  \newunit\newblock
  \usebibmacro{addendum+pubstate}%
  \setunit{\bibpagerefpunct}\newblock
  \usebibmacro{pageref}%
  \usebibmacro{finentry}}
  
\renewbibmacro*{volume+number+eid}{%
  \printfield{volume}%
  \setunit*{\addnbspace}
  \printfield{number}%
  \setunit{\addcomma\space}%
  \printfield{eid}}
  
\AtEveryBibitem{\clearlist{organization}}
\AtEveryBibitem{\clearfield{issue}}
\AtEveryBibitem{\clearfield{number}}
\AtEveryBibitem{\clearfield{note}}

\usepackage{setspace}
\begin{document}
\onehalfspacing
\title{An in-plane photoelectric effect in two-dimensional electron systems for terahertz detection}
\date{}
\author{Wladislaw Michailow$^1$, Peter Spencer$^1$, Nikita W. Almond$^1$, Stephen J. Kindness$^1$,\\Robert Wallis$^1$, Thomas A. Mitchell$^1$, Riccardo Degl'Innocenti$^2$,\\Sergey A. Mikhailov$^3$, Harvey E. Beere$^1$, and David A. Ritchie$^1$\\\\\small
1 -- Cavendish Laboratory, University of Cambridge, J J Thomson Avenue, Cambridge CB3 0HE, United Kingdom\\\small
2 -- Department of Engineering, University of Lancaster, Bailrigg, Lancaster LA1 4YW, United Kingdom\\\small
3 -- Institute of Physics, University of Augsburg, D-86135 Augsburg, Germany}

\maketitle

\large 
\begin{abstract}
\large
The photoelectric effect consists in the photoexcitation of electrons above a potential barrier at a material interface and is exploited for photodetection over a wide frequency range. This three-dimensional process has an inherent inefficiency: photoexcited electrons gain momenta predominantly parallel to the interface, while to leave the material they have to move perpendicular to it. 
Here, we report on the discovery of an \textit{in-plane photoelectric effect} occurring within a two-dimensional electron gas. In this purely quantum-mechanical, scattering-free process, photoelectron momenta are perfectly aligned with the desired direction of motion. The ``work function'' is artificially created and tunable \textit{in-situ}. The phenomenon is utilized to build a direct terahertz detector, which yields a giant zero-bias photoresponse that exceeds the predictions by known mechanisms by more than 10-fold. This new aspect of light-matter interaction in two-dimensional systems paves the way towards a new class of highly efficient photodetectors covering the entire terahertz range.

\end{abstract}

\section{Introduction}

A key milestone in the history of quantum mechanics was the discovery of the photoelectric effect \cite{Lenard1902}: a photon excites an electron from the surface of a metal, if its quantized energy $\hbar\omega$ is larger than the metal's workfunction \cite{Einstein1905}. Apart from this \textit{external} photoelectric effect, at lower frequencies (far infrared -- visible) photoexcitation of electrons into higher-energy quantum states within a semiconductor can lead to a photoresponse of the material, which is an \textit{internal} photoeffect.

The photoemission from a material in three dimensions is a complicated process consisting of several steps \cite{Berglund1964,Broudy1971}: first photons excite electrons, these then move and scatter within the solid, and finally leave into the collecting medium provided they have sufficient energy and momentum \textit{perpendicular} to the interface. This leads to its inherent inefficiency: the photoexcited electrons gain predominantly a momentum $\vec p$ parallel to the electric field \cite{Sass1975} and thus \textit{parallel} to the surface under normal incidence of radiation. Electrons can obtain a perpendicular momentum component, needed to escape from the material, under oblique incidence of p-polarised light \cite{Broudy1971, Girardeau1993,Hechenblaikner2014}, or by scattering processes that randomise the direction of motion of photoexcited electrons \cite{Berglund1964}. But scattering reduces their energy, diminishes the efficiency, and limits the intrinsic response time of the effect to scattering times.

Being a fundamental physical phenomenon converting light (photons) to electricity, the photoelectric effect can be exploited for detection of electromagnetic radiation over a wide range of frequencies. In the mid- and far infrared ranges, it has been used in internal photoemissive detectors \cite{Perera1992,Perera1995,Shen2000,Lao2014}. However, the responsivity of homojunction and heterojunction internal photoemission detectors rapidly falls off below 5\,THz, essentially vanishing below $\sim$ 2.5\,THz \cite{Matsik2003,Perera2008,Bai2018}. Furthermore, these three-dimensional, bulk photoemissive detectors are photoconductive and thus require an applied external bias \cite{Shen1997,Haller1979,Stacey1992,Bai2018}.

In this paper, we report on a giant, direct THz photoresponse in a two-dimensional electron gas (2DEG). In our devices, we cover a 2DEG in a semiconductor heterostructure with two gates which simultaneously serve as a terahertz (THz) antenna. By applying different voltages to the gates, we create a potential step for electrons within the plane of the 2D channel. Under THz irradiation, which is normally incident on the semiconductor surface, electrons are then instantaneously excited above the potential step, which leads to a giant photocurrent and photovoltage response under zero source-drain bias. The electric field of the wave and the desired direction of photoelectron momentum are perfectly aligned in the same direction, parallel to the surface. We will call this the \textit{in-plane photoelectric effect} (not to be confused with the lateral photoelectric effect \cite{Zalinge2004}, an unrelated phenomenon). It is a direct quantum-mechanical, scattering-free process, which thus does not have an intrinsic response time limit due to scattering. The height of the potential step, an analog of the work function in our device, is electrically tunable \text{in situ} by applying different voltages to the gates.

Previously, THz detection in 2DEGs and structures based on those has been realised using various mechanisms, in III-V and silicon semiconductors \cite{Molenkamp1990,Janssen1994,Kouwenhoven1994a,Kouwenhoven1994b, Drexler1995,Verghese1995, Lu1998,Knap2002,Song2010, Nadar2010,Knap2011,AlHadi2012,Levin2015, Otsuji2015,Otteneder2018}, as well as, more recently, also in novel 2D materials \cite{Zak2014,Viti2016, DeglInnocenti2016}. The discovered effect we report is more than 10-fold as strong as predicted by the classical plasmonic mixing \cite{Dyakonov1996,Knap2002} in two-dimensional systems, and cannot be explained by other mechanisms such as photon-assisted tunneling \cite{Tkachenko2015, Otteneder2018}, bolometric, electron heating \cite{Song2010, Levin2015}, or thermoelectric effects \cite{Molenkamp1990}. We develop a quantum theory of the in-plane photoelectric effect which describes the experimental findings very well. We show that the discovered phenomenon is ideally suited for highly efficient photodetection across the entire terahertz range.

\section{Results}

\subsection{Physics of the phenomenon}

\begin{figure}[!b]
\centering
\includegraphics[width=\textwidth]{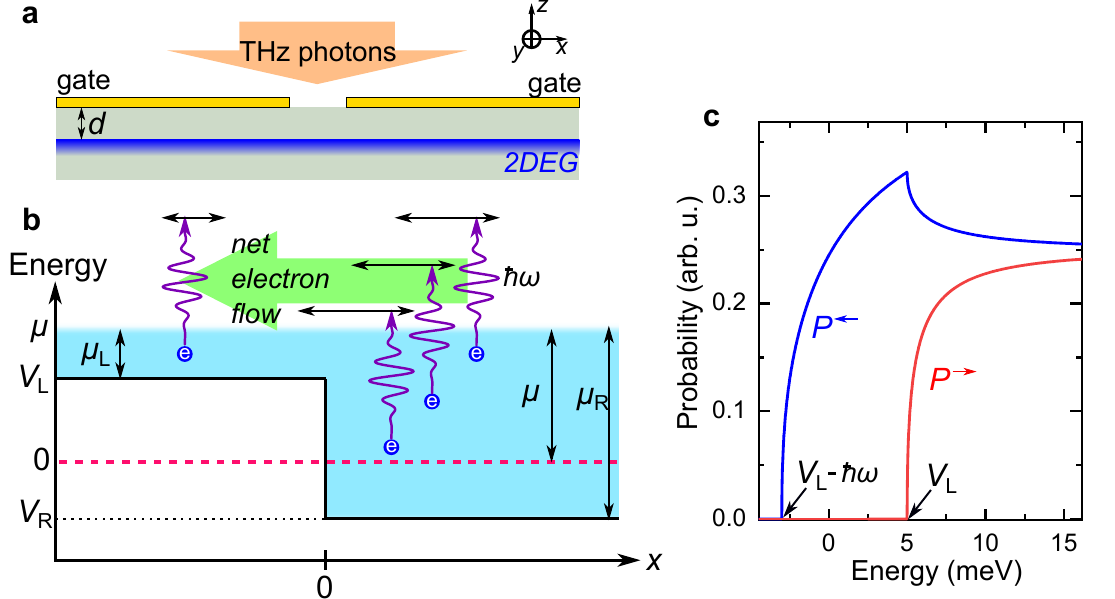}
\caption{\label{physprinciple}\textbf{Physical principle of the detection mechanism based on the in-plane photoelectric effect.} (a) Side view of the device: a semiconductor heterostructure with a 2D electron gas is covered by two gates and irradiated by electromagnetic radiation. (b) Energy diagram of the device. $\mu$ is the equilibrium chemical potential, $V_\mathrm L>0$ and $V_\mathrm R<0$ are the conduction band edges in the left and right parts of the device. A potential step of value $V_\mathrm L-V_\mathrm R$ is artificially created by applying different voltages to the gates. This gives rise to different electron densities and different chemical potentials, $\mu_\mathrm L=\mu-V_\mathrm L$ and $\mu_\mathrm R=\mu-V_\mathrm R$, in the left and right areas of the 2D electron gas. Incident electromagnetic radiation on a potential step excites electrons on both sides of the barrier, resulting in a net electron flow onto the step. (c) Energy dependence of the probability to absorb a photon and pass the step for electrons moving from right to left (blue curve) and from left to right (red curve), see Methods. The curves are plotted for $\hbar\omega=8$\,meV, $V_\mathrm{L}=5$\,meV, and $V_\mathrm{R}=-5$\,meV.
}
\end{figure}

Consider a uniform 2DEG covered by two gates, see Fig. \ref{physprinciple} (a). In equilibrium, the bottom of the conduction band (dashed purple line) and the chemical potential $\mu$ are position-independent. By applying two different voltages on the gates, $U_\mathrm{G,L}$ and $U_\mathrm{G,R}$, we create an \textit{artificial, gate voltage tunable potential step} for electrons moving in the 2DEG. This shifts the bottom of the equilibrium conduction band under the left and right gates by $V_\mathrm L$ and $V_\mathrm R$, respectively, resulting in the local chemical potentials $\mu_\mathrm L$ and $\mu_\mathrm R$, see Fig. \ref{physprinciple} (b). Without any incident radiation, the net particle flow vanishes: electrons with energies $E<V_\mathrm L$ do not contribute to the current since they are reflected backwards at the step, while for $E>V_\mathrm L$ the current flow cancels out due to equal reflection and transmission probabilities regardless of the direction of electron motion.

Once the potential step in the gap between the gates is exposed to incident electromagnetic radiation of frequency $f = \omega/(2\pi)$, electrons can absorb a photon of energy $\hbar\omega$. Those with energies $E<V_\mathrm L$ on the right side, that previously could not overcome the step, are now able to do so by absorbing incident photons. For electrons with energies $V_\mathrm L<E<\mu$, photon absorption will lead to a higher probability to move to the left (onto the step) than to the right, see Fig. \ref{physprinciple} (c). As a result, the radiation induces a net particle flow from the higher-density region to the lower-density region. This can be understood as a photoelectric effect occuring within the plane of the 2DEG.

This phenomenon can be used for detection of far-infrared and terahertz radiation. We utilize the in-plane photoelectric effect for THz detection in a 2DEG covered by two gates that simultaneously serve as a THz antenna.

\subsection{Samples and setup}

\begin{figure}[ht]
\centering
\includegraphics[width=\textwidth]{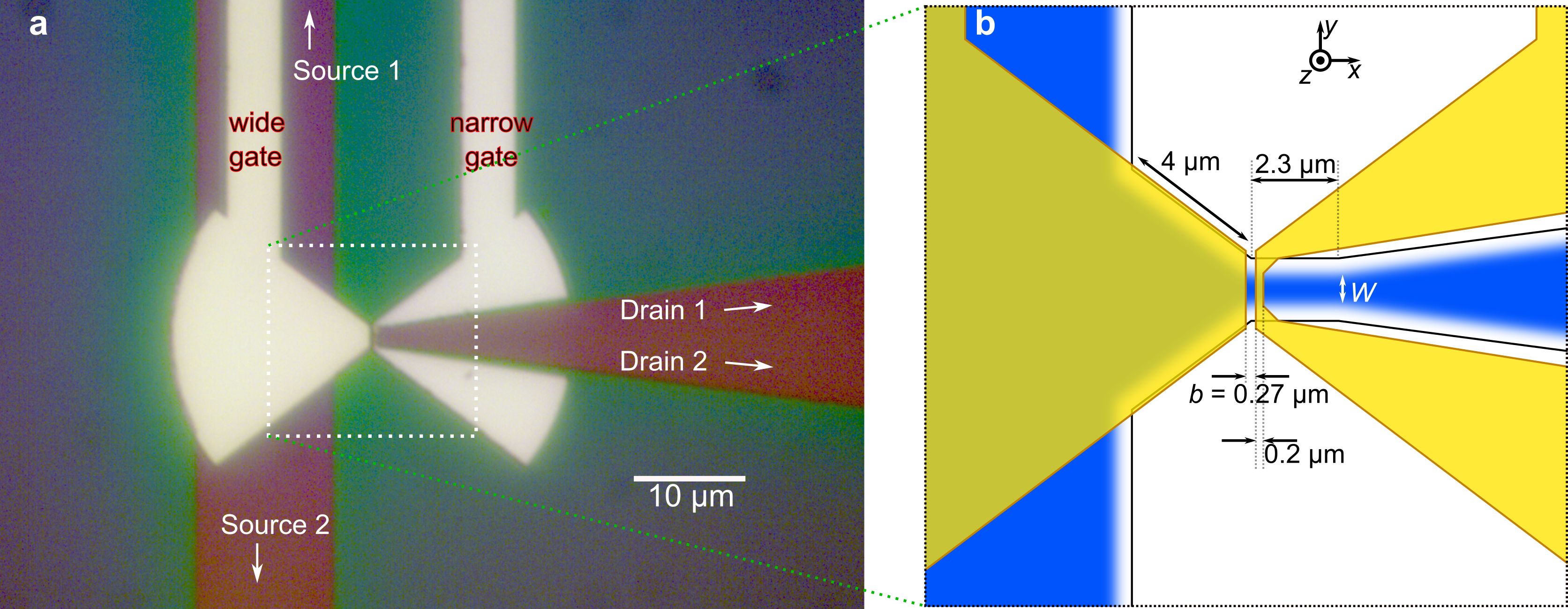}
\caption{\label{samplesetup}\textbf{Sample view.} (a) Optical microscope image of the sample: the wide and narrow gates are made in the form of a bow-tie antenna; the 2DEG is contacted by two source and two drain Ohmic contacts; (b) schematic illustration of the device: the black shape shows the contour of the mesa, while the blue area shows the 2DEG. Its blurred edges account for uncertainty in the amount of edge depletion; $W\sim 0.5 - 1$\textmu{}m. The metallised gates are shown semitransparent in yellow.}
\end{figure}

Our samples, see Fig. \ref{samplesetup} (a), are fabricated from a GaAs-Al$_{0.33}$Ga$_{0.67}$As heterostructure grown by molecular beam epitaxy (see details in Methods section). The 2DEG in the wafer material, $d\approx 90$\,nm below the surface, exhibits an electron density of $3.3\cdot 10^{11}/$cm$^2$ and a mobility of $2.1\cdot 10^6$ cm$^2$/(Vs) at 1.5\,K, corresponding to a scattering time of $\tau=80$\,ps and a mean free path of 20\,\textmu{}m. Narrow channels were created by mesa etching. The active part of the device, see Fig. \ref{samplesetup} (b), is covered by a bow-tie antenna that couples THz radiation to the 2DEG. The right part of the antenna, the ``narrow gate'', is split into two halves that are joined together via a 200\,nm narrow bridge, Fig. \ref{samplesetup} (b). Both wings of the antenna serve as gates in our device.
The 2DEG has four Ohmic contacts, labelled Source 1 or 2, and Drain 1 or 2, which are at least 100 \textmu{}m away from the center of the antenna to avoid their influence on the active part of the device.

The devices were cooled to 9\,K $(k_\mathrm BT\approx 0.78\,\mathrm{meV})$ in a liquid helium continuous flow cryostat. A copper waveguide system delivered 1.9\,THz $(\hbar\omega\approx 7.9\,\mathrm{meV})$ radiation from a quantum cascade laser (QCL) source \cite{Faist1994,Koehler2002} to the samples. The QCL is electrically modulated with a modulation frequency of $f_\mathrm{mod}=781$\,Hz and a duty cycle of 2.14\,\%. The time-averaged intensity incident on the sample is (for details see Methods section) $\langle I\rangle =I_0\cdot 2.14\%=6.25$\,\textmu{}W/mm$^2$. Within a pulse, the intensity is $I_0=0.29$\,mW/mm$^2$.

\subsection{Photoresponse and device parameters}

\begin{figure}[ht]
\includegraphics[width=\textwidth]{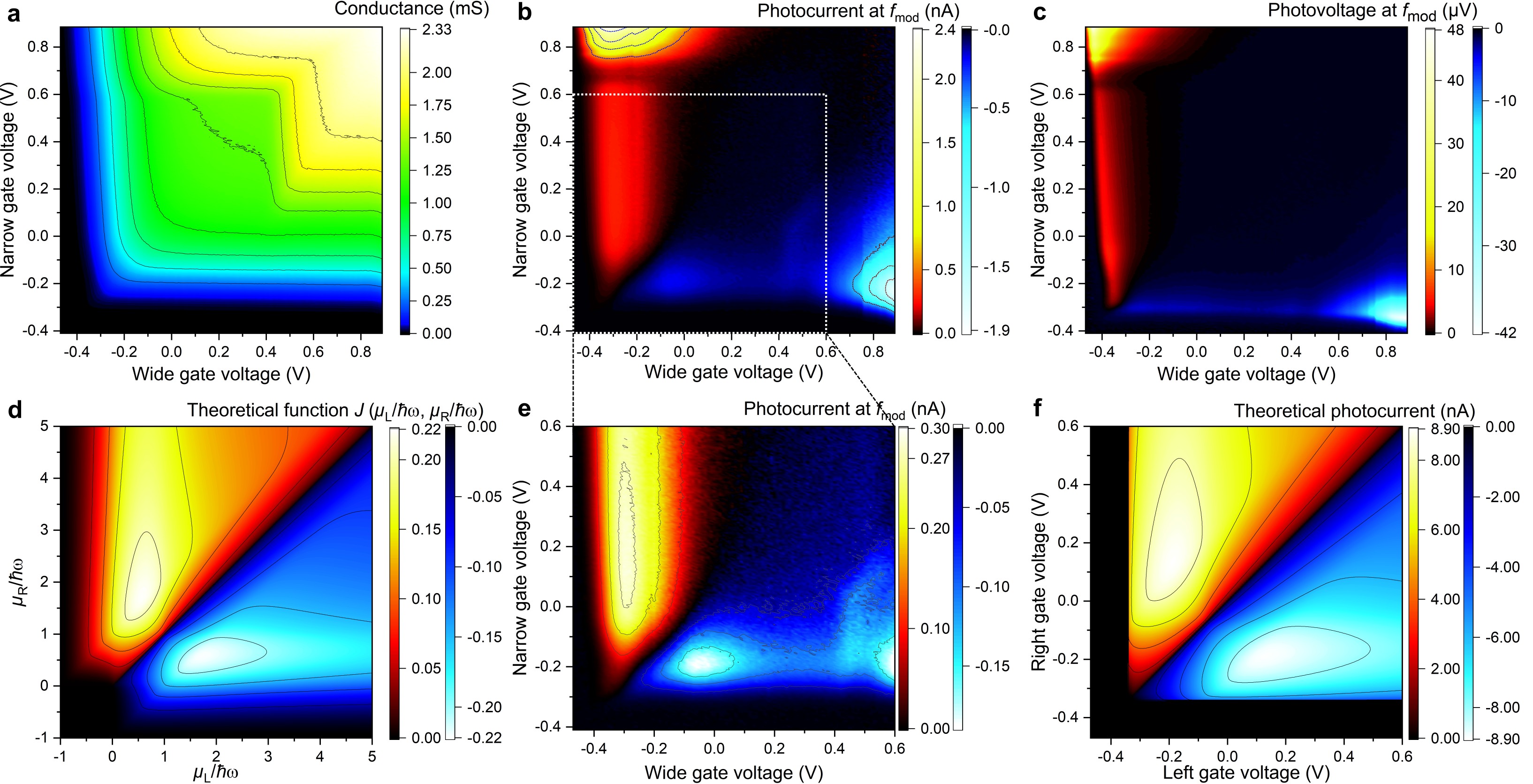}
\caption{\textbf{Gate-voltage dependent measured device parameters and supporting theoretical calculations.} \textit{Experimental results:} (a)-(c), (e): 2D maps as a function of wide gate voltage (horizontal axis) and narrow gate voltage (vertical axis); (a): 4-wire conductance of the device. (b): Photocurrent, with a detailed plot of the region $U_\mathrm{G,wide}, U_\mathrm{G, narrow}<0.6\,$V in (e); (c): photovoltage. The photoresponse to incident THz radiation is measured as time-averaged value at the modulation frequency of the quantum cascade laser, corresponding to the intensity $\langle I\rangle =6.25$\,\textmu{}W/mm$^2$. \textit{Theoretical calculations:} (d) Function $J(\mu_\mathrm L/\hbar\omega; \mu_\mathrm R/\hbar\omega)$ from Eq. (\ref{photocurrent}) as a function of the ratios of left and right chemical potentials to the incident photon energy $\hbar\omega$. (f) Theoretically expected photocurrent in the region $U_\mathrm{G,wide}, U_\mathrm{G, narrow}<0.6\,$V shows good agreement with experimental data in (e). Negative and positive values in (b)-(f) are presented in separate color bars. Contour line values are indicated on the colorbars.\label{exp-theor}}
\end{figure}

Fig. \ref{exp-theor}\,(a) illustrates the conductance of the sample, measured in a 4-terminal configuration, as a function of the two gate voltages in a 2D colormap. At zero gate voltages, the 4-wire channel resistance is $R_0\approx 1.1$\,k\textOmega{}. The conductance vanishes at the left and at the bottom of the map, which corresponds to the pinch-off of the channel using the wide or narrow gate, respectively. The threshold voltage determined from the wide gate is $U_\mathrm{th}\approx -0.35$\,V; here the 4-wire channel resistance is 16 times higher than $R_0$. The channel is switched off for $U_\mathrm{G,wide}<-0.44$\,V, where the resistance exceeds 1\,M\textOmega{}. $U_\mathrm{th}$ corresponds to an equilibrium electron density of $n_\mathrm s=-\epsilon_0\epsilon_\mathrm rU_\mathrm{th}/(ed)\approx 2.7\cdot 10^{11}/\mathrm{cm}^2$ and an equilibrium chemical potential of $\mu = \pi\hbar^2 n/m_\mathrm{eff}\approx 9.6\,\mathrm{meV}$, with $\epsilon_\mathrm r\approx 12.6$ and the effective electron mass $m_\mathrm{eff}=0.067m_\mathrm e$ in GaAs \cite{Blakemore1982}.

Under incident THz radiation, the sample generates a photoresponse. The induced current under  zero source-drain bias in a short-circuit measurement, the photocurrent, is shown in Fig. \ref{exp-theor}\,(b). The photovoltage acquired in an open-circuit measurement is depicted in Fig. \ref{exp-theor}\,(c). The measurement is done using a lock-in amplifier with the QCL modulation frequency $f_\mathrm{mod}$ as reference. The red and blue color schemes indicate areas where the response is positive or negative, respectively.

The photoresponse arises predominantly in the left or bottom areas in the 2D maps Fig. \ref{exp-theor}\,(b) and (c), where either of the narrow or wide gate voltages are negative. 
Here, one of the gates depletes the 2DEG, which gives rise to a potential step in the channel. The sign of the photoresponse is always such that the THz-induced electron flow moves onto the potential step, from the higher density region to the lower density region. The photoresponse is mostly anti-symmetric with respect to the diagonal line $U_\mathrm{G, narrow}=U_\mathrm{G, wide}$. Two regimes can be identified: (i) $U_\mathrm{G,wide}$ and $U_\mathrm{G,narrow} \lesssim 0.6$ V, shown in detail in Fig. \ref{exp-theor} (e), and (ii) $U_\mathrm{G,wide}$ or $U_\mathrm{G,narrow} \gtrsim 0.6$ V, where the photoresponse increases strongly in the top-left and bottom-right corners. The latter regime corresponds to strong asymmetry of the gates, where one of the gates operates in the depletion regime with a negative voltage applied with respect to the source Ohmic contact, while the other gate has a strong positive voltage applied and operates in the enhancement mode.

A further increase in the gate voltage, beyond $U_\mathrm{G, narrow}>0.885\,$V and $U_\mathrm{G, wide}>0.75\,$V, leads to the onset of a rapidly growing gate leakage. In this regime, undesirable for detector operation, the sample becomes unstable and noisy.
Blocking and unblocking the incident THz radiation did not result in any change (within measurement accuracy of 10\,pA) of the gate currents. This excludes gate currents as a possible origin of the photoresponse, e.\,g. due to Schottky barrier rectification.

A negative voltage on the narrow gate creates a barrier of 0.2\,\textmu{}m width, whereas on the wide gate it creates a $\gtrsim 4$\,\textmu{}m wide barrier (Fig. \ref{samplesetup} (b)). A remarkable result is that the photoresponse changes by less than 30\,\% (Fig. \ref{exp-theor} (b), (c) top-left vs. bottom-right), in spite of a more than 20-fold difference in barrier width. This rules out a tunneling origin of the effect.

\begin{figure}[th]
\centering
\includegraphics[width=\textwidth]{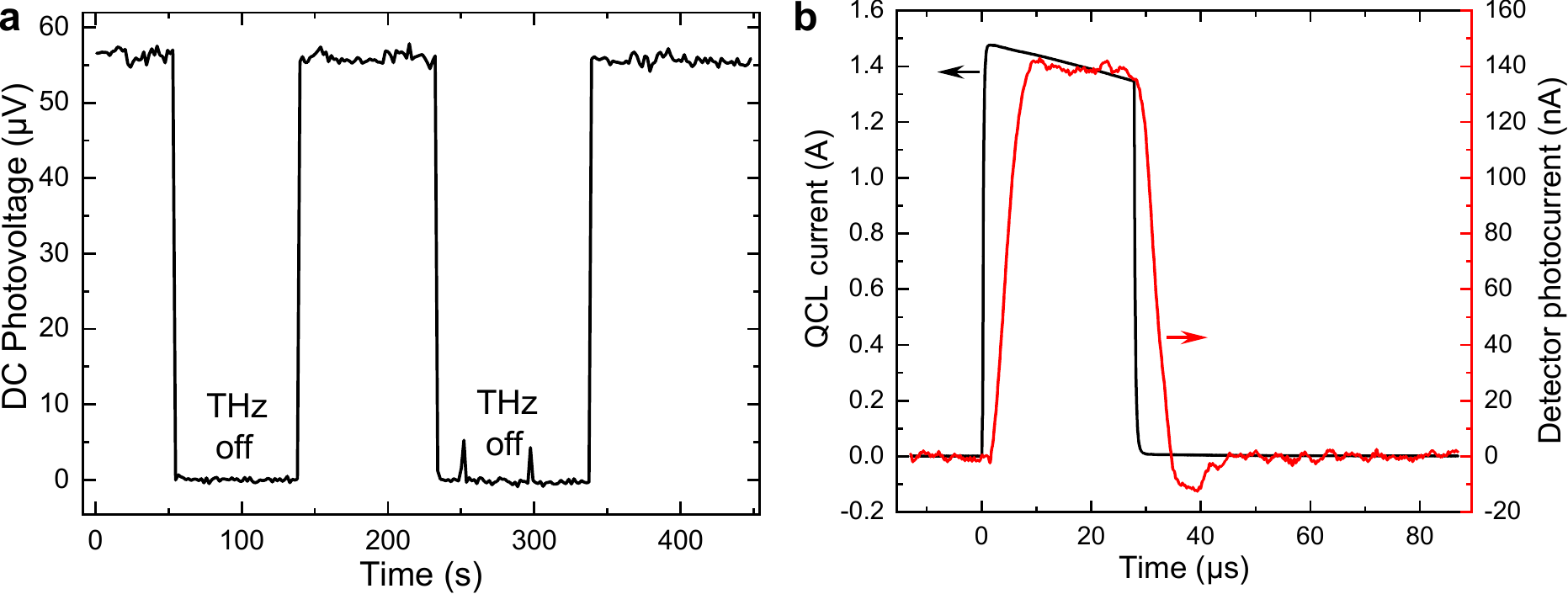}
\caption{\label{QTSD-usage}\textbf{THz detection with the presented device.} (a) DC source-drain voltage response in photovoltage readout mode, when the THz waveguide is repeatedly mechanically blocked and unblocked. When the THz waveguide is open, QCL emission with 2.14\% duty cycle is incident on the sample. A maximum time-averaged DC photovoltage of 56\,\textmu{}V is induced by the electrically modulated QCL emission. (b) Sample response to incident THz pulses. Left axis: QCL current, right axis: measured detector current. A photocurrent of $\sim$ 142\,nA is observed during a THz pulse emitted by the QCL.}
\end{figure}

The best photovoltage performance is achieved at the point $(U_\mathrm{G, wide}, U_\mathrm{G, narrow})=(-0.45\mathrm V,0.885\mathrm V)$, Fig. \ref{exp-theor} (c), where the resistance is $\sim 70$\,k\textOmega{}. At this point, Fig. \ref{QTSD-usage} (a) shows the source-drain voltage measured using a DC voltmeter, while the THz radiation is mechanically blocked and unblocked. The clear DC response demonstrates that the device can operate as a direct THz detector. The DC photovoltage response is 56\,\textmu{}V and the absorption cross section of the antenna is (22.5\textmu{}m)$^2$ (obtained from numerical simulations). Together with the incident intensity $\langle I\rangle$ this gives the photovoltage responsivity of\,17.6 kV/W. With 0.5\,\textmu{}V root-mean-square voltage noise within 50\,Hz bandwidth, the noise-equivalent power can be estimated to be 4\,pW/$\sqrt{\text{Hz}}$, which is still limited by the noise of the experimental setup.

\begin{figure}[ht]
\centering
\includegraphics[width=\textwidth]{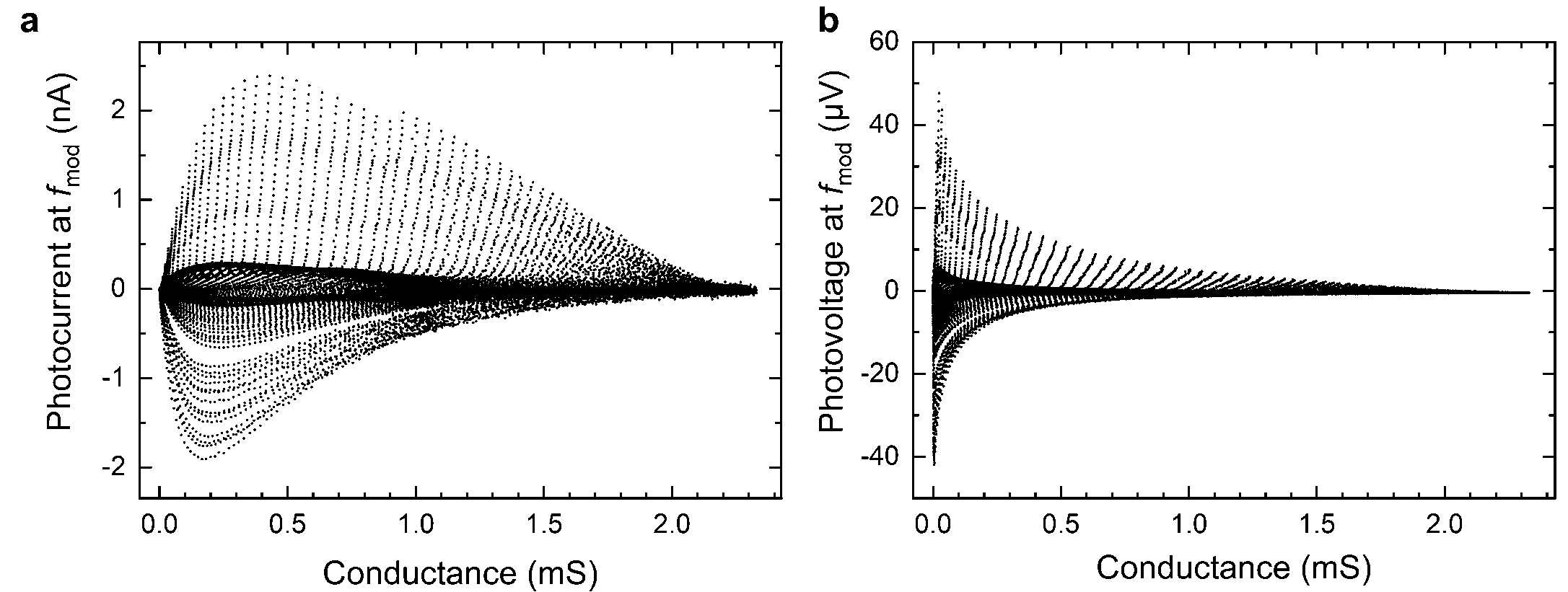}
\caption{\label{responsivityranges} \textbf{(a) Photocurrent and (b) photovoltage as a function of 4-wire sample conductance.} Each data point corresponds to a certain value of the wide and narrow gate voltages. The area enclosed by the data points indicates the available range of responsivity and resistance. For impedance matching to external circuits, the wide and narrow gate voltages are chosen at the intersection of the desired photoresponse and output impedance.}
\end{figure}

Optimal photocurrent readout is achieved at the point $(U_\mathrm{G, wide}, U_\mathrm{G, narrow})=(-0.32\mathrm V,0.885\mathrm V)$ in Fig. \ref{exp-theor}\,(b), where the low output impedance of the device, $\sim 2.6$\,k\textOmega{}, allows for fast detection of the photocurrent. Notably, the best photocurrent is thus achieved when the device is far from pinch-off: the channel resistance is merely 2.4 times higher than at zero gate voltages. Fig. \ref{QTSD-usage}\,(b) shows the response to incident THz pulses from the QCL on a microsecond time scale under these conditions.
With a photocurrent of $\sim$142\,nA during the pulse and a pulse intensity of $I_0\approx$ 0.29 mW/mm$^2$, the photocurrent responsivity is 0.96 A/W. 

A key advantage of the two gates is the independent tuning of output impedance and responsivity. This enables impedance matching to external circuitry -- a capability which is inherently built-in in the device design. The target working point for a desired photocurrent response and output impedance, i.e. the two gate voltages, is to be determined from the intersection of equi-photocurrent contours in Fig. \ref{exp-theor}\,(b) with equi-conductance contours in Fig. \ref{exp-theor}\,(a). This also means that detection sensitivity can be switched on or off with the gate voltages, while maintaining the device's impedance, thus making it appear the same in the external circuit. The available responsivity and output impedance ranges can be found in Fig. \ref{responsivityranges}.

\subsection{Theory} 

We consider a narrow 2DEG under two semi-infinite gates. The gate voltages $U_\mathrm{G,L}$ and $U_\mathrm{G,R}$ induce a step-like potential $V_0(x)=V_\mathrm L-(V_\mathrm L-V_\mathrm R)\Theta(x)$, see Fig. \ref{physprinciple}. 
Under THz irradiation, an additional time-dependent potential $V_1(x,t)=(eE_\mathrm{ac}b/2)\mathrm{sign}(x) \cos(\omega t)$ is added to $V_0(x)$. $E_\mathrm{ac}$ is the amplitude of the THz electric field in x-direction acting on the 2DEG at the step, and $E_\mathrm{ac}b$ the potential difference it induces. Electrons are assumed to pass the gap $b$ between the antenna wings ballistically, i.e. to have a mean free path exceeding $b=0.27$\,\textmu{}m (this condition is satisfied in the experiment). In the y-direction, an infinite potential well of width $W$ is assumed ($W\approx 0.7$ \textmu{}m, see Methods), so that the ground-state energy of the transverse motion is $E_W=\pi^2\hbar^2/(2mW^2)\approx 12$ \textmu{}eV. 

To describe the photoresponse of the system, we solve the time-dependent Schr\"odinger equation in the first-order perturbation theory, treating $V_1$ as a perturbation, see Methods. The resulting expression for the photocurrent is
\be 
I_\mathrm{ph}=ef
\left(\frac{eE_\mathrm{ac}b}{\hbar\omega}\right)^2
\left[ 2 \sqrt{\frac{\hbar\omega} {E_W}}  J\left(\frac{\mu_\mathrm L}{\hbar\omega},\frac{\mu_\mathrm R}{\hbar\omega}\right)\right].
\label{photocurrent}
\ee
 The photocurrent's dependence on the chemical potentials is described by a universal dimensionless function $J$ that depends on the two dimensionless parameters $\mu_\mathrm L/(\hbar\omega)$ and $\mu_\mathrm R/(\hbar\omega)$ and is antisymmetric, see Fig. \ref{exp-theor} (d). 
The gates are pinched off at $\mu_\mathrm {L/R}=0$. In the area with $\mu_\mathrm {L}<0$ and $\mu_\mathrm {R}<0$, where there are no electrons in both parts of the sample, the photocurrent vanishes.
When only one of the chemical potentials is negative, down to $-\hbar\omega$, the function $J$ is non-zero: photoexcited electrons are still able to jump onto the step when $\mu < V_\mathrm{L/R} < \mu+\hbar\omega$.

The maximum of the function $J(\mu_\mathrm L/\hbar\omega; \mu_\mathrm R/\hbar\omega)$ equals $0.221$ and is located at $\mu_\mathrm L=0.57\hbar\omega$; $\mu_\mathrm R= 1.73\hbar\omega$. Notably, this corresponds to the open regime far
away from pinch-off, where in both parts of the system, $x>0$ and $x<0$, electrons form a degenerate Fermi gas with chemical potentials well above the conduction band edges, as illustrated in Fig. \ref{physprinciple} (b).
This relationship between the radiation frequency and the chemical potentials allows to determine the optimal frequency range of the detection mechanism. Since typical electron concentrations $n=m_\mathrm{eff}\mu/(\pi\hbar^2)$ in 2DEGs range from $\lesssim 10^{10}$/cm$^2$ \cite{Hirayama1998} to $\gtrsim 4\cdot 10^{12}$/cm$^2$ \cite{Mokerov1999}, optimal photodetection can be achieved between 0.15 -- 20\,THz. The in-plane photoelectric effect is thus ideally suited for detection of radiation in the entire THz range.

In Fig. \ref{exp-theor}\,(f) we plot the (time-averaged) theoretical photocurrent expected under the experimental conditions, using the relationship between chemical potentials and gate voltages as described in the Methods section. 
There is a good qualitative agreement with the experimental photocurrent for $U_\mathrm{G, wide}, U_\mathrm{G, narrow}<0.6$\,V in Fig. \ref{exp-theor}\,(e). In this regime the measured photocurrent peaks at $(U_\mathrm{G, wide}, U_\mathrm{G, narrow})=(-0.29\mathrm V,0.15\mathrm V)$ in Fig. \ref{exp-theor}\,(e), with a channel resistance of 3.7\,k\textOmega{}. The position of maximum theoretical photocurrent in Fig. \ref{exp-theor}\,(f) is close to the experimental peak position. The match between theory and experiment is best when considering the top-left regions (where $I_\mathrm{ph}>0$). This is sensible since it corresponds to a barrier under the wide gate, which is a better approximation of a semi-infinite potential step as considered in the theory, than the 200\,nm narrow gate. In spite of the simplicity of the model, all key features are thus reproduced. A better agreement between theory and experiment can be achieved by considering a more detailed model of the potential, also taking into account the ungated region between the wide and narrow gates.

The strong increase in photocurrent under maximum asymmetry in the top-left and bottom-right regions in Fig. \ref{exp-theor}\,(b), which is not covered in the theoretical framework, is likely related to an effect of the device geometry, such as a widening of the conducting channel, since edge depletion effects start to diminish at strongly positive gate voltages. An increase in channel width, which is $\sim 0.5 - 1$\textmu{}m in equilibrium as indicated in Fig. \ref{samplesetup}\,(b), would give rise to an increased conductance in Fig. \ref{exp-theor}\,(a), as the bottleneck in the form of the narrow channel is widened, and also to an enhanced photocurrent $I_\mathrm{ph}\sim W$, which increases with the channel width.

Quantitatively, the theoretically calculated photocurrent of $8.9$\,nA exceeds the experimentally measured value. This is in contrast to other possible interpretations, discussed in the next section, which predict a lower value than measured. The experimental photocurrent in the region $U_\mathrm{G, wide}, U_\mathrm{G, narrow}<0.6$\,V, Fig. \ref{exp-theor}\,(e) is $0.3$\,nA, and the overall maximum measured photocurrent is $2.4$\,nA, Fig. \ref{exp-theor}\,(b). This indicates that the experimentally observed signal can be further increased by optimizing the device design to reduce channel and contact resistances which diminish the measured current.

\subsection{Discussion}

We now check whether other detection mechanisms could explain the observed effect. Bolometric mechanisms commonly give rise to photoconductance, rather than a zero-bias photocurrent or -voltage \cite{Janssen1994,Song2010,Levin2015}, and dominate in the pinch-off regime, where conduction is sensitive to small temperature changes \cite{Song2010}. In contrast, in our case, the photocurrent is maximal in the open regime. In the case of a broken symmetry, a photocurrent may be observed, e.\,g. as a (photo-)thermoelectric effect \cite{Molenkamp1990}. However, the prerequisite for the applicability of such interpretations is local electron thermalisation. But in our case, the mean free path $\lambda_\mathrm e$ of electrons is very large: 20\,\textmu{}m in the wafer material, and even considering a reduction of $\lambda_\mathrm e$ within the channel due to the edge roughness, it still exceeds $b=0.27\,$\textmu{}m, the distance between the gates, which is the relevant length scale. Thus, electrons pass the region between the gates ballistically. This rules out local electron thermalisation, and mechanisms relying on it.

On the other hand, electron ``heating'' can also occur in a collisionless way, in the sense that the incident radiation increases the mean electron energy, as in Ref. \cite{Levin2015}. The increase in the mean energy of the electrons can be estimated as $e^2E_\mathrm{ac}^2/(4m_\mathrm{eff}\omega^2)\approx 0.46$\,\textmu{}eV$\ll k_\mathrm BT\ll\mu$, which is a negligible amount (for the maximum electric field in x-direction of $E_\mathrm{ac}=100$\,V/cm during a THz pulse with intensity $I_0$, extracted from numerical simulations of the antenna amplification).

Another mechanism to consider is photon-assisted tunneling \cite{Verghese1995,Tkachenko2015,Otteneder2018}: photons incident on a thin barrier in the tunneling regime can give rise to photoconductance. However in the present experiment, the size of the barriers created by the gates is macroscopically large ($\sim 4$\,\textmu{}m in the case of the wide gate), and the photocurrent dominates in the open regime, with a degenerate electron gas. This rules out tunneling-related effects such as photon-assisted tunneling.

On the other hand, photon-assisted tunneling across a thin barrier can be understood as the difference of two photocurrents generated at the two edges of the barrier, which is considered as two mirrored potential steps. The total photocurrent will vanish at zero bias due to the inherent symmetry of the barrier, and a photoresponse will only be observed if the symmetry is intentionally broken by an applied voltage across the barrier. This explains why the photoresponse of thin tunneling systems such as quantum point contacts presents itself predominantly as photoconductance \cite{Song2010,Levin2015,Otteneder2018}, rather than zero-bias photocurrent or photovoltage \cite{Janssen1994}. In our device, the translation symmetry is broken by the gate-induced potential step without any external source-drain bias, which is advantageous for noise performance. We implement the potential step using a macroscopically wide barrier and irradiate it on one side only. Thus we recover the giant photocurrent signal that tends to remain hidden in quantum point contacts.

Many THz detection experiments \cite{Lu1998,Knap2002,Lisauskas2009,Nadar2010,Knap2011,Boppel2012,AlHadi2012,Zak2014,Viti2016} were interpreted in terms of the plasma-wave mixing theory \cite{Dyakonov1996}, a mechanism which at lower frequencies corresponds to distributed resistive mixing or resistive self-mixing \cite{Lisauskas2009, Boppel2012}. 
Let us estimate the absolute value predicted by the plasmonic mixing mechanism. In our experiment, the THz gate-to-channel voltage amplitude $U_\mathrm{ac}$ equals 0.18\,mV for the average incident power according to numerical antenna simulations.
The sample can be modelled as a back-to-back series connection of two 2DEGs with different densities. The source-drain distance is much longer than the plasmon decay length, and $\omega\tau\gg 1$. Under these conditions, the plasmonic mixing theory \cite{Dyakonov1996} predicts a photovoltage of 0.2\,\textmu{}V at the point of maximum experimental photocurrent in the $U_\mathrm G<0.6\,$V area, $(U_\mathrm{G, wide}, U_\mathrm{G, narrow})=(-0.29\mathrm V,0.15\mathrm V)$, where the measured photovoltage is 1.8\,\textmu{}V, i.\,e. considerably higher. 
Furthermore, the maximum possible photovoltage for non-resonant detection is $\approx eU_\mathrm{ac}^2/(4\eta k_\mathrm B T)\sim 1$\,\textmu{}V, according to Ref. \cite{Knap2002}, with $\eta\gtrsim 10$ in GaAs based FETs at low temperatures \cite{Knap2002}.
This is smaller than the experimental photovoltage of 5.6\,\textmu{}V in the area $U_\mathrm G<0.6\,$V, and more than an order of magnitude smaller than the maximal DC photovoltage of  56\,\textmu{}V at strong gate asymmetry as seen in Fig. 3\,(a). In addition, even the qualitative form of the photoresponse cannot be explained: plasmonic mixing would result in a monotonous increase of the photocurrent the further apart the gate voltages are, and the broad maximum of the photocurrent observed experimentally would not be reproduced. Thus, other possible interpretations, as discussed above, cannot explain the effect.

In conclusion, we have demonstrated a fundamental physical phenomenon -- the in-plane photoelectric effect. It has a number of advantages as compared to the conventional photoelectric effect. It is a purely quantum-mechanical, scattering-free phenomenon which has no intrinsic response time limit. The maximum photocurrent is obtained when the chemical potential lies above the potential step in both parts of the 2DEG, i.e., the 2DEG is well conducting. Such a situation has not been implemented so far, and would not work in a 3D case. No dc source-drain bias is required, whereas in 3D photoemissive detectors usually a bias is applied to extract the electrons. The effect is more than 10 times stronger than any mechanism previously considered in 2D electron systems. The potential step is artificially created and tunable by gate voltages; in contrast, in 3D photoemissive systems the workfunction is fixed by the material parameters. The effect is ideally suited for utilization across the whole THz range. As an inherent effect of 2D systems, it can be utilized in 2DEGs on the basis of III-V materials, silicon, as well as the novel 2D layered, graphene-related materials. 

The detection principle based on the discovered phenomenon paves the way to a new class of \textit{photo-electric tunable-step} (PETS) THz detectors. The presence of two gates allows independent tuning of output impedance and responsivity of the device, which facilitates integration into external circuits. The dual-gated, antenna-coupled device architecture will advance the development of fast and large-scale integrated THz detectors and focal plane arrays. Integrating PETS detectors with quantum cascade lasers, as is commonly done with photodiodes and optical laser chips, will be a valuable advancement towards market-ready THz QCL-based emitters.

\section{Methods}

\subsection{Sample fabrication}

The heterostructure used was deposited on a GaAs semi-insulating substrate. The sequence grown by molecular beam epitaxy was, after a 1\,\textmu{}m undoped GaAs buffer layer, 40 nm undoped Al$_{0.33}$Ga$_{0.67}$As; 40 nm n-doped Al$_{0.33}$Ga$_{0.67}$As with $10^{18}$/cm$^3$ doping density of Si donors; 10 nm GaAs cap.
To process the sample, electron beam lithography is used to define an etch mask using ma-N 2410 resist onto the as-grown wafer. At this stage, the sample is etched to form a 100 nm high mesa with the narrow channel and areas for contact pads. The Ohmic source and drain contacts to the channel are processed using optical lithography by annealing an AuGeNi eutectic alloy.
Next, electron beam lithography with poly(methyl methacrylate) resist is used to define a TiAu gate structure on top of the device which creates a bow-tie antenna that focuses the radiation to the channel. Optical lithography
is used for creation of large-area TiAu bond pads. Finally, the sample is encapsulated in Shipley S1805 resist for surface protection.

\subsection{Edge depletion}

The lithographically defined channel width is 2.0\,\textmu m. Calibration of the mesa etching has shown that sideways etching of 165\,nm occurs on average on either side of the mesa, so the actual mesa width is approximately 1.67 \textmu{}m. Measurements on other samples created by this process have shown that the lateral edge depletion of the 2DEG is between 0.34 \textmu{}m and 0.58 \textmu{}m. Taking this into account, the actual width $W$ of the 2DEG in the channel, is between $0.5$ and $1$ \textmu{}m, and is assumed to be 0.7\,\textmu{}m in the theory.

\subsection{Electrical measurement setup}
The conductance in Fig. \ref{exp-theor} (a) was calculated as the ratio of current to voltage when the sample was driven with a 1.6 mV sine wave at 86.5 Hz. The current is passed through Source 1 and Drain 1, the voltage is measured from Source 2 to Drain 2 in Fig. \ref{exp-theor} (a). 

\subsection{Determination of the power density at the sample}

To illuminate the sample with THz radiation, we have constructed a setup coupling together two liquid helium continuous flow cryostats, one with a 1.9\,THz single-plasmon QCL, and the other with the sample, using a copper waveguide. This waveguided setup allows quantitative determination of the THz intensity at the sample space. Before measurements, the QCL cryostat is first aligned so as to induce maximum photoresponse of the sample. 
After the measurements, the sample is warmed up and removed, while the QCL is kept lasing, and its temperature (18\,K) and alignment are maintained.
The total transmitted power at the sample space is then determined using a Thomas Keating absolute power meter, while the intensity distribution is measured using a Golay cell with a 1 mm aperture on a setup with motorised x, y scanning stages. We thus find the time-averaged intensity $\langle I \rangle=6.25$\,\textmu{}W/mm$^2$ at the sample space at the mode profile peak, where the intensity distribution reaches its maximum value.

\subsection{Photocurrent calculation}

The dynamics of photoexcited electrons is determined by the time-dependent Schrödinger equation:
\be i\hbar \frac{\partial\Psi}{\partial t}=
-\frac{\hbar^2}{2m}\frac{\partial^2\Psi}{\partial x^2} -\frac{\hbar^2}{2m}\frac{\partial^2\Psi}{\partial y^2}+V_0(x)\Psi+V_1(x,t)\Psi\,.
\label{SE}
\ee
In the transverse, y-direction we assume infinite potential walls at $y=0$ and $y=W$, which gives 
quantization energies of $E_Wn^2$, where $E_W=\pi^2\hbar^2/(2mW^2)\approx 12$ \textmu{}eV and $n$ is the subband index.

Solving Eq. (\ref{SE}) with $V_1$ as perturbation, we calculate the wavefunctions in zero and  first order, and from them, the particle flows $j_{E,n}^{0\leftarrow}$, $j_{E,n}^{0\rightarrow}$ and $j_{E,n}^{+\leftarrow}$, $j_{E,n}^{+\rightarrow}$. Here, $j_{E,n}^{0\leftarrow}$ is the flow of particles with energy $E$ incident onto the step from the right, and $j_{E,n}^{+\leftarrow}$ is the flow of particles with energy $E+\hbar\omega$ having passed the step and moving to the left (the definitions for ``$\rightarrow$'' are similar). Their ratio leads to the first-order transmission coefficients of particles having absorbed a THz quantum $\hbar\omega$, defined as $T_{E,n}^{+\rightarrow}=j_{E,n}^{+\rightarrow}/j_{E,n}^{0\rightarrow}$ for positive, and $T_{E,n}^{+\leftarrow}=j_{E,n}^{+\leftarrow}/j_{E,n}^{0\leftarrow}$ for negative $x$-direction of motion. 
Thus calculated transmission coefficients can be written in the form
$T_{E,n}^{+\rightleftarrows }=\alpha P^{\rightleftarrows }(E-E_Wn^2,\hbar\omega,V_\mathrm{L},V_\mathrm{R})\,,$
where $\alpha=\left(eE_\mathrm{ac}b/\hbar\omega\right)^2$ and the functions $P^{\rightleftarrows }(E,\hbar\omega,V_\mathrm{L},V_\mathrm{R})$ are illustrated in Fig. \ref{physprinciple} (c) for $\hbar\omega=8$\,meV, $V_\mathrm{L}=5$\,meV, and $V_\mathrm{R}=-5$\,meV. As seen from the plot, $T_{E,n}^{+\leftarrow}>T_{E,n}^{+\rightarrow}$ for all energies provided that $V_\mathrm L>V_\mathrm R$, i.\,e. the particles flow onto the step. The photocurrent $I_\mathrm{ph}$ is then calculated as
\be I_\mathrm{ph}=-\frac{e}{\pi\hbar} \sum_{n=1}^\infty \int_{-\infty}^\infty \mathrm dE \left(T_{E,n}^{+\rightarrow} - 
T_{E,n}^{+\leftarrow} \right)f(E,\mu,T)\left(1-f(E+ \hbar\omega,\mu,T)\right)
\label{photocurrent-general}
\ee
where the factors containing the Fermi distribution function $f(E,\mu,T)$ take into account the occupation of electron states and $\mu$ is the chemical potential of the unbiased system. 
Eq. (\ref{photocurrent}) is obtained from Eq. (\ref{photocurrent-general}) at temperature $T=0$.

The perturbation theory parameter $\alpha$ should be $\ll 1$ for the applicability of the theory. With the incident intensity of $I_0$ during a THz pulse, the electric field in x-direction acting on the 2DEG between the gates is $E_\mathrm{ac,x}\approx 100$\,V/cm according to the numerical antenna simulations in Comsol Multiphysics. This gives $eE_\mathrm{ac}b\approx 2.7$\,meV and $\alpha\approx 0.12\ll 1$, which justifies the perturbative approach and neglection of higher-order contributions to the photocurrent.

In order to transform the dependencies on chemical potentials in Eq. (\ref{photocurrent}) and Fig. \ref{exp-theor} (d) into gate voltage dependencies as shown in Fig. \ref{exp-theor} (f), the connection between external gate potentials $U_\mathrm{G, L}$, $U_\mathrm{G, R}$ and chemical potentials $\mu_\mathrm L=\mu-V_\mathrm L$,  $\mu_\mathrm R=\mu-V_\mathrm R$ is required. This relationship can be obtained from the Poisson equation, taking into account the screening of the external potential created by the gates, 

\be
eU_\mathrm{G, L}=(\mu_\mathrm L-\mu)+\frac{4d}{a_\mathrm B^*}k_\mathrm B T\ln\left(\frac{1+e^{\mu_\mathrm L/(k_\mathrm BT)}}{1+e^{\mu/(k_\mathrm BT)}}\right)\,,
\ee
and analogously, for $U_\mathrm{G, R}$ and $\mu_\mathrm R$. Here $\mu=\pi\hbar^2 n_0/m_\mathrm{eff}$ is the equilibrium chemical potential, $n_0$ the equilibrium 2D electron density, and $a_\mathrm B^*=4\pi\epsilon_0\epsilon_\mathrm r\hbar^2/(m_\mathrm{eff}e^2)$ the effective Bohr radius in GaAs.

\section{Acknowledgements}

W.M. thanks the George and Lilian Schiff Studentship of the University of Cambridge for financial support and is grateful for the Honorary Vice-Chancellor’s Award of the Cambridge Trust. S.A.M. acknowledges funding from the European Union's Horizon 2020 research and innovation programme Graphene Core 3 under Grant Agreement No. 881603. R.D. acknowledges support from the EPSRC (Grant No. EP/S019383/1). The authors acknowledge EPSRC funding within the Hyperterahertz grant, number EP/P021859/1, and thank Binbin Wei, Yuqing Wu, Ateeq Nasir, Ben Ramsay, and Antonio Rubino for helpful advice, as well as Abbie Lowe and Jonathan Griffiths for help with electron beam lithography. The authors express special thanks to Joanna Waldie for advice on sample fabrication and measurements and comments on the manuscript.

\section{Author information}

\subsection{Contributions}

W.M. conceived the device concept, designed, simulated, fabricated, and measured the samples, built the waveguide-coupled terahertz setup and wrote the paper. P.S. advised on instrumental techniques, helped and supported in the lab. N.W.A. built the motorised Golay cell scanning setup. S.J.K. advised on Comsol numerical simulations. R.W. advised on terahertz measurement techniques. T.A.M. did the lithographic electron beam exposure. R.D. participated in discussions of experimental findings. S.A.M. provided quantitative theory and corresponding contribution to the paper. H.E.B. grew the wafer material and advised. D.A.R. enabled and supervised the research and advised. All authors discussed the results and the manuscript.

\subsection{Corresponding author}

Correspondence to W. Michailow (wm297@cam.ac.uk).

\printbibliography

\end{document}